\documentclass[conference]{IEEEtran}
\IEEEoverridecommandlockouts
\usepackage{amsmath,amssymb,amsfonts}
\usepackage{algorithmic}
\usepackage{graphicx}
\usepackage{textcomp}
\usepackage{xcolor}
\usepackage[hidelinks]{hyperref}
\usepackage{booktabs}
\usepackage{todonotes}
\usepackage{array}
\usepackage{amsmath}
\usepackage{booktabs}
\usepackage{comment}
\usepackage{subfig}
\usepackage{arydshln}
\usepackage{multirow}

\usepackage[backend=biber,sorting=none, style=ieee]{biblatex} %
\def\BibTeX{{\rm B\kern-.05em{\sc i\kern-.025em b}\kern-.08em
    T\kern-.1667em\lower.7ex\hbox{E}\kern-.125emX}}

\newcommand{\frameworkname}{\textsc{EEGain}}

\begin{document}

\title{Evaluation in EEG Emotion Recognition: State-of-the-Art Review and Unified Framework}

\author{
Natia Kukhilava, Tatia Tsmindashvili, Rapael Kalandadze, Anchit Gupta,
Sofio Katamadze,\\Fran\c{c}ois Br\'emond, Laura M. Ferrari, Philipp Müller, and Benedikt Emanuel Wirth
\thanks{Natia Kukhilava, Tatia Tsmindashvili, Rapael Kalandadze, and Sofio Katamadze are at the Muskhelishvili Institute of Computational Mathematics, Georgian Technical University, Tbilisi, Georgia}
\thanks{Anchit Gupta, Philipp Müller, and Benedikt Emanuel Wirth are at the German Research Center for Artificial Intelligence, Saarbrücken, Germany}
\thanks{Laura M. Ferrari is at The Biorobotics Institute, Scuola Superiore Sant’Anna, Pontedera, Italy; INRIA, Universite Côte d’Azur, Sophia Antipolis, France}
\thanks{Fran\c{c}ois Br\'emond is with INRIA, Universite Côte d’Azur, Sophia Antipolis, France}
\thanks{This research was funded by the European Union under Horizon Europe project GAIN (GA \#101078950) and partially funded by PNRR - Investment 1.5 Ecosystems of Innovation, ProjectTuscany Health Ecosystem (THE), Spoke 3 “Advanced technologies, methods, materials and health analytics” CUP:I53C22000780001.}
}

\maketitle

\begin{abstract}

Electroencephalography-based Emotion Recognition (EEG-ER) has become a growing research area in recent years. Analyzing 216 papers published between 2018 and 2023, we uncover that the field lacks a unified evaluation protocol, which is essential to fairly define the state of the art, compare new approaches and to track the field's progress. We report the main inconsistencies between the used evaluation protocols, which are related to ground truth definition, evaluation metric selection, data splitting types (e.g., subject-dependent or subject-independent) and the use of different datasets.
Capitalizing on this state-of-the-art research, we propose a unified evaluation protocol, \frameworkname{} (\url{https://github.com/EmotionLab/EEGain}), which enables an easy and efficient evaluation of new methods and datasets. \frameworkname{} is a novel open source software framework, offering the capability to compare---and thus define---state-of-the-art results.
\frameworkname{} includes standardized methods for data pre-processing, data splitting, evaluation metrics, and the ability to load the six most relevant datasets (i.e., AMIGOS, DEAP, DREAMER, MAHNOB-HCI, SEED, SEED-IV) in EEG-ER with only a single line of code. %
In addition, we have assessed and validated \frameworkname{} using these six datasets on the four most common publicly available methods (EEGNet, DeepConvNet, ShallowConvNet, TSception). 
This is a significant step to make research on EEG-ER more reproducible and comparable, thereby accelerating the overall progress of the field.

\end{abstract}


\section{Introduction}

Electroencephalography based Emotion Recognition (EEG-ER) has become popular in the past years~\cite{zhang2020emotion, maithri2022automated, torres2020eeg, https://doi.org/10.1155/2020/8875426}, as the brain is regarded as the site where emotional activities are evoked.
EEG-ER promises to go beyond observation of outward behaviours, such as facial expressions~\cite{videoemorec}, speech~\cite{speechemorec}, or body movements~\cite{muller2015emotion}, which are subject to cultural norms~\cite{matsumoto1990cultural, safdar2009variations} and potentially under users' conscious control~\cite{mcconatha1999primary, gross1998emerging}. 
EEG provides a means to directly access the internal state of users~\cite{shu2018review} through a non-invasive interface. EEG is recorded by means of multiple skin-contact electrodes placed in specific locations that measure the electrical potential on the scalp, which reflects the underlying neural activity~\cite{buzsaki2006rhythms, light2010electroencephalography}.

Despite its potential, extracting information on emotional states from the EEG is challenging, for the following reasons.
The signal is noisy~\cite{subha2010eeg, annurev:/content/journals/10.1146/annurev.bioeng.5.040202.121601}, it is prone to artifacts~\cite{jiang2019removal}, and in addition to emotional states EEG reflects many other psychological processes such as perception, attention, and memory~\cite{s23146434}.
The datasets are small, comprising less than 100 participants~\cite{amigos, deap, dreamer, mahnob, seeda, seedb, seed-iv}. The complex process of placing electrodes on participants' scalps leads to large expenses in dataset recording.
The EEG exhibits high variability between participants~\cite{9748967, Zhao_Yan_Lu_2021}, and even between recordings of the same participant at different points in time~\cite{APICELLA2024128354}.
There are significant deviations between recording set-ups used in different datasets (e.g., the number and kind---wet or dry--- of electrodes, emotion elicitation procedures, etc.).

The dominant approaches to handle this challenging signal consist of a sequence of signal pre-processing techniques and machine learning (ML) methods
\cite{cimtay2020investigating,iyer2023cnn,li2022sstd,wang2023self}.
The pre-processing typically comprises filtering, downsampling, windowing, artifact removal and normalization. While in the ML process the main aspects are related to the selection of the dataset, the train-test splitting, the ground truth splitting, and the the evaluation metrics.

While other reviews in the field of EEG-ER exist \cite{multiemorecreview, APICELLA2024128354, 7946165, https://doi.org/10.1155/2020/8875426, 9492294}, they  primarily describe the datasets and methods used, but do not address a significant aspect, the evaluation process. This paper aims to fill this gap by investigating all the pipeline contents, from the data to the evaluation metrics. 
We conducted a thorough review of 216 papers on EEG-ER, published between 2018 and 2023, identifying \textit{five key issues} that prevent a clear identification of state-of-the-art approaches.
First, different studies use different \textit{datasets} in order to train and test their approaches. Second, different studies employ different pre-processing steps to the raw EEG signal. Third, the \textit{data splitting}, that is, how the data is divided into train, validation, and test subsets varies between studies. Fourth, different studies use different \textit{ground truth} definitions by applying inconsistent binarization thresholds for continuous valence and arousal ratings.
Finally, the employed \textit{evaluation metrics} vary substantially between different studies.

Taken together, these five points are a barrier to reproducibility and fair comparison, thereby severely limiting progress in the field.
Based on the thorough review reported here, we propose  \frameworkname{}, a unified evaluation framework for EEG-ER.
\frameworkname{} provides easy-to-use and standardized solutions for the whole training and evaluation pipeline.
It provides functions to load the most popular EEG-ER datasets into a common format and to perform the necessary pre-processing. In detail, we provide code to load the following datasets, AMIGOS~\cite{amigos}, DEAP~\cite{deap}, DREAMER~\cite{dreamer}, MAHNOB-HCI~\cite{mahnob}, SEED~\cite{seeda, seedb}, and SEED-IV~\cite{seed-iv}.
Furthermore, \frameworkname{} allows to train and evaluate several recent state-of-the-art EEG-ER methods, as well as custom approaches that can be implemented by the user.
The models supported out-of-the-box by \frameworkname\ are EEGNet \cite{lawhern2018eegnet}, DeepConvNet \cite{schirrmeister2017deep}, ShallowConvNet \cite{schirrmeister2017deep} and TSception \cite{tsception2022}.
Finally, prediction results can be evaluated with metrics such as F1, Precision, Recall, Accuracy, Kappa, MCC, and confusion matrices with just a few lines of code. 
Using this framework, researchers can easily evaluate their methods in a standardized, comparable way. To summarize, our contributions are three-fold.
\begin{itemize}

\item[1)] We review the state of the art in EEG-ER analyzing and identifying the shortcomings of the adopted evaluation protocols.

\item[2)] We propose \frameworkname{}, a framework that provides easy-to-use functions for standardized and comparable pre-processing, training, and evaluation of EEG-ER methods.

\item[3)] We compare four recent methods on six different datasets with the unified  \frameworkname{} framework.

\end{itemize}

As such, our work lays the foundation for reliable evaluation and quantifiable progress in EEG-ER.

\section{Background}

In the following, we describe five key characteristics which serve as a background for our literature review.
These are: (A) the \hyperref[sec:Datasets]{datasets} chosen as input, (B) the \hyperref[sec:Pre-processing]{pre-processing} methods applied to the raw data, (C) the \hyperref[sec:Data Splitting]{data splitting} into train and test data, (D) the \hyperref[sec:Ground Truth]{ground truth} definition for dimensional and categorical emotion models, and (E) the \hyperref[sec:Evaluation Metrics]{evaluation metrics}.

\subsection{Datasets}
\label{sec:Datasets}
Selecting datasets is the first step in any kind of machine learning research. The available datasets in the EEG-ER domain are diverse, with variations that are mostly related to the set-ups employed during EEG recording~\cite{amigos, deap, dreamer, mahnob, seeda, seedb, seed-iv}.
The recording procedure usually follows a specific scheme. First, the recruited subjects, which are typically healthy volunteers, are equipped with the devices required for the recording (i.e. the EEG headset, plus other optional sensors for the recording of electrocardiography, electrodermal activity, etc.). Subsequently, subjects are placed in front of a screen and different emotions are elicited trial-by-trial. At the end of each trial, the subjects are asked to rate their emotions using a self-assessment scale, as detailed in the \hyperref[sec:GT]{Ground Truth} Section. Typically, consecutive trials are separated by a cool-down time to avoid contamination of one trial's signal by the previous trial.

One primary source of variance are the stimuli used to elicit emotional responses. These stimuli can be images, videos, or other sensory triggers. This variance is substantial as different stimuli can elicit various intensities and types of emotional responses. An additional concern is the challenge of ascertaining that the stimuli effectively elicit the intended emotions consistently across different subjects. Any potential inconsistency in this regard can affect the reliability of the data, raising questions about whether the recorded emotional responses accurately reflect the desired emotional states.

Another factor of variation is the trial duration, which can vary between 1 and 4 minutes. For longer experiments, it is important to note that emotional responses typically last only a few seconds.
As a result, the predominant mood captured in recordings that extend to several minutes is often a neutral state \cite{s22134939, s23187853}.

The employed EEG headset is another source of variance. The market offers a spectrum of devices, from low-priced, consumer-oriented to sophisticated headsets for research purposes. High-end headsets typically provide superior-quality recordings with less noise, whereas consumer-grade devices, though more accessible, usually provide poorer signal quality. Furthermore, the electrode type employed during recording (e.g., gel vs. dry) significantly affects the quality of the data. Gel-based headsets are generally acknowledged for providing lower impedances and, consequently, recordings with higher quality than their dry counterparts~\cite{6679670, 10.3389/fnins.2024.1326139}.

Additionally, the structure of the data collection process varies across datasets. In general, recordings consist of multiple trials per participant. However, the scheduling of these trials can differ – in some datasets, they were conducted on a single day~\cite{mahnob, deap, amigos, dreamer}, while in others they were spread across multiple days in separate sessions~\cite{seeda, seedb, seed-iv}. These temporal discrepancies are non-trivial as EEG recordings are sensitive to a multitude of factors, including the individual's sleep patterns and other day-to-day physiological variations.

All of these dataset-related sources of variance impede fair comparisons between different emotion-recognition approaches that are trained and evaluated on different datasets.

\subsection{Pre-processing}
\label{sec:Pre-processing}

Data pre-processing is used to mitigate the noisy nature of EEG. %
In the following, we introduce the most used pre-processing techniques for EEG-ER (see~\cite{RN1005, RN1006} for more detailed introductory explanations on pre-processing techniques commonly used in EEG research).

\subsubsection{Notch Filter}
A Notch filter is a type of filter that removes a specific narrow band of frequencies and passes frequencies outside this band. It is usually used to remove the line frequency, which is linked to the electrical standards of the country where the recording takes place. For instance, in Europe it is 50 Hz, while in the USA it is 60 Hz. 

\subsubsection{Band-Pass Filter}
A Band-pass filter is a type of filter that passes frequencies inside a specified band and removes frequencies below or above this range. Band-pass filtering helps isolate selected frequency components, while attenuating environmental noise and other signal components. 
Typically a 4th order band-pass Butterworth filter is used in EEG processing. 
Moreover, band-pass filtering can be used to narrow the frequency range of interest. For instance, alpha (8-13 Hz), beta (13-30 Hz), and delta (0.1-4 Hz) frequency bands have been extracted and linked to emotional states \cite{ZHANG2023104157}. It has been shown that selecting certain bands can help to enhance the signal-to-noise ratio and improve the accuracy of emotion recognition algorithms~\cite{8703559, 10.3389/fnins.2022.985709}.

\subsubsection{Artifact Removal}
EEG datasets are often contaminated with various artifacts, such as eye blinks, muscle and heart activity, and other sources of noise. Artifact removal techniques involve the use of algorithms or manual inspection to identify and eliminate these unwanted components from the EEG data. Independent Component Analysis (ICA), Principal Component Analysis (PCA) and wavelet denoising are the most frequently used techniques to  remove the artifacts from the EEG data \cite{JUNG_MAKEIG_HUMPHRIES_LEE_McKEOWN_IRAGUI_SEJNOWSKI_2000, TONIN2024578, CASTELLANOS2006300}. 

\subsubsection{Downsampling}
Emotion recognition datasets have different sampling rates (data points per second). While in basic neuroscientific research, EEG data are typically recorded and analyzed at 500 Hz~\cite{RN907, RN272} to 1000 Hz~\cite{RN393, RN932}, the sampling rates at which datasets in the field of EEG-ER can be downloaded are generally lower, ranging between 128 Hz~\cite{amigos, dreamer} to 512 Hz~\cite{deap}.
To compare the results across multiple datasets, EEG signals need to be downsampled to the same minimum sampling rate.
Notably, downsampling decreases the dataset size, as well as the overall training time.

\subsubsection{Windowing}
Windowing involves segmenting continuous EEG data into shorter time intervals. A common window size of 2 seconds is used for ER tasks~\cite{s22134939, 7835688}. This process effectively partitions the continuous EEG signal into distinct segments, each associated with the ground truth label given to the full session. Typically, consecutive windows overlap ensuring that the information contained within each window is not isolated but shared between neighboring segments, enhancing the continuity and coherence of the data analysis process.
One significant advantage of employing windowing is that after this process, input size is fixed, whereas in many datasets the full session length can vary, typically between 1-4 minutes, as detailed in the \hyperref[sec:Datasets]{Datasets} section. Another advantage is the ability to manage memory constraints and computational efficiency. When dealing with entire sessions, high-ended hardware is needed. 
The main disadvantages of windowing are linked to the loss of time information and increased ground truth related weakness. Ground truth weakness or noise refers to inaccuracies present in dataset labels, which can arise in scenarios involving self-assessment where individuals evaluate their own emotions. This subjectivity means that identical responses from two different individuals may not necessarily refer to the same emotion, thereby introducing noise into the dataset. %

\subsubsection{Normalization}
Normalization is primarily used to ensure that data from different subjects, recording sessions, or sensors are on a consistent scale~\cite{10.3389/fnins.2021.626277, s23187749}. %
Normalization can involve various methods, such as z-score normalization or min-max scaling.

\subsection{Data Splitting}
\label{sec:Data Splitting}

For any ML approach, it is crucial to separate data into training and test splits to avoid overfitting and ascertain appropriate generalization of the trained model to new data.
Two types of data splitting are common in the field of EEG-ER: subject-independent and subject-dependent splits. 

\subsubsection{Subject-independent}

In the subject-independent approach, the training and test sets consist of recordings from different individuals.
As a result, models need to be able to generalize across individuals, which is a challenging task considering the large inter-subject variability in EEG recordings~\cite{9748967, Zhao_Yan_Lu_2021}.
Subject-independent train/test splits can be either realized by a fixed split, or in a cross-validation procedure.
With small datasets, the cross-validation approach has advantages, as training set sizes are maximized and evaluation is performed on all available subjects.
Subject-independent cross-validation can be performed in two ways.
The \textit{leave-one-subject-out} (LOSO) procedure uses a single subject's data as the test set in each cross-validation iteration.
More specifically, with $n$ subjects, this results in $n$ cross-validation iterations, requiring $n$ models to be trained.
While this maximizes training set sizes, it is also computationally expensive due to the large numbers of models to be trained. In the field of EEG-ER, LOSO is the most common form of subject-independent splitting
The \textit{leave-k-subjects-out} (LkSO) approach uses $k$ subjects as test sets in each iteration.
The consequence are fewer cross-validation iterations, leading to smaller computational demands.
On the downside, the training set sizes are smaller compared to the LOSO technique.

\subsubsection{Subject-dependent}

In the subject-dependent approach, the goal of generalization is restricted to a specific individual.
Recordings of each subject are divided into two parts: one for training and the other for testing. 
As the dataset size is an even larger concern than in the subject independent scenario, cross-validation is preferable over fixed train-test splits.
Analogously to the subject-independent framework, two main approaches are used. 
The \textit{leave-one-trial-out} (LOTO) approach, where a single trial (e.g. defined by the presentation of an emotional video clip) is used as the test set in each iteration of the cross-validation. The remaining trials of the subject are used for training. This is the approach requiring the least generalization capabilities from models, as subject-dependent factors remain a constant.
The \textit{leave-k-trials-out} (LkTO) approach decreases the number of cross-validation iterations (and the training set sizes) by using several trials as test sets in each cross-validation loop.
In certain datasets (e.g., SEED, SEED-IV), multiple recordings were obtained from each person across several days. 
Each day's recording is referred to as a session. 
The \textit{leave-one-session-out} technique involves selecting one session per person for the test set while utilizing the remaining sessions for training.
It poses a greater generalization challenge compared to the LOTO approach due to the temporal variability of the EEG signal~\cite{APICELLA2024128354}.

\subsection{Ground Truth}
\label{sec:Ground Truth}

Emotions are a complex concept and different ways of describing emotions exists, most notably the dimensional and categorical emotion models.

\subsubsection{Dimensional Emotion Models}
\begin{figure}[htbp]
    \centering
    \subfloat[The circumplex model of emotion. The model assumes that each emotion can be described by two dimensions, valence and arousal.]{
        \includegraphics[width=0.45\textwidth]{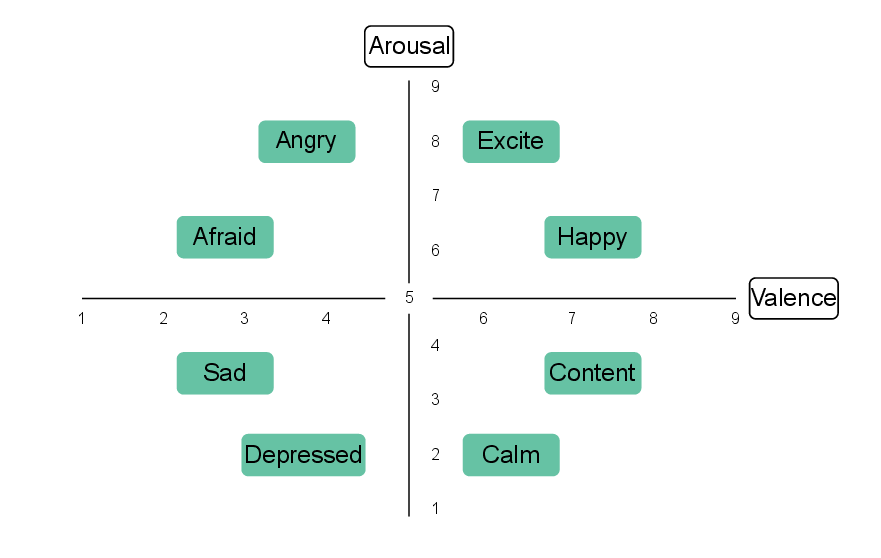}
        \label{fig:valence-arousal scale}
    }
    \hfill
    \subfloat[The Self-Assessment Manikin (SAM). This self-assessment scale allows subjects to intuitively judge their current emotional state.]{
        \includegraphics[width=0.45\textwidth]{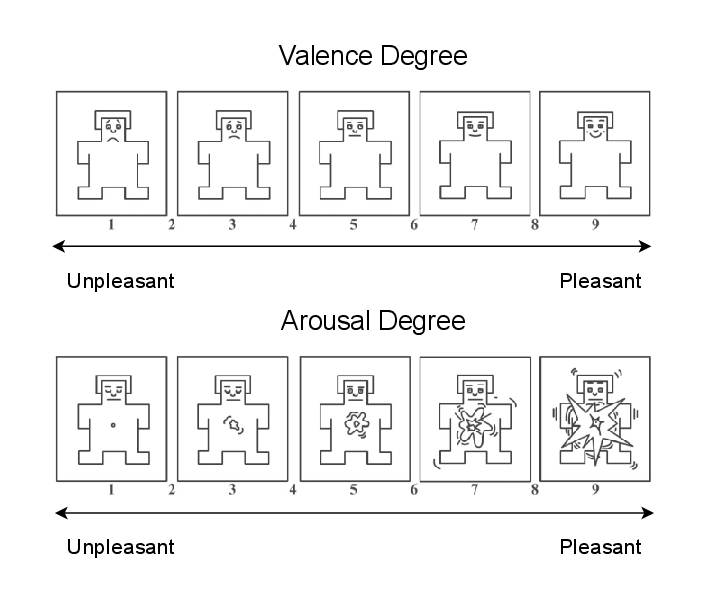}
        \label{fig:er-self-assessment}
    }
    \caption{(a) The circumplex model of emotion, (b) The Self-Assessment Manikin (SAM).}
    \label{fig:combined-figures}
\end{figure}

Dimensional emotion models represent an emotional state as a point in a continuous space of two or more dimensions.
The most common model is the circumplex model, which consists of a \textit{valence} and an \textit{arousal} dimension~\cite{russell1980circumplex}.
Valence represents the subjective pleasantness of an emotion and ranges from negative (i.e., unpleasant) to positive (i.e., pleasant). Arousal represents the level of affective activation from sleep to excitement (see Fig.~\ref{fig:valence-arousal scale}).
The PAD model \cite{mehrabian1996pleasure} additionally incorporates a \textit{dominance} dimension which reflects the level of control or influence a person feels in a particular situation, ranging from submissive to dominant. 
Later work has argued that the fourth dimension of \textit{unpredictability} representing the degree to which a person is surprised should be added~\cite{fontaine2007world}, but this model is only rarely used in affective computing research ~\cite{muller2015emotion, mahnob}
In practice, most EEG-ER approaches use only valence and arousal to predict a person's emotional state~\cite{8703559, s21103414, 10.1145/1500879.1500888}.
Typically, these two dimensions are measured with a self-assesment scale that participants complete at the end of each session. This scale is based on a pictogram, either continuous or discrete, ranging from 1-5 or 1-9 (see Fig.~\ref{fig:er-self-assessment} for an example).

To simplify the emotion prediction task, researchers often binarize the problem by defining high (1) vs. low (0) arousal and high (1) vs. low (0) valence classes.
As described below, an open point in this regard is the specific definition of the threshold. 
Exploring multi-class or regression models could offer a more detailed understanding of emotional states, preserving the emotional nuances. The main limitation to this approach, however, is the small size of the available datasets.

\subsubsection{Categorical Emotion Models}
The categorical view on emotions ~\cite{BasicEmotions} claims that a small set of universal basic emotions exist. According to the most common assumption, the set comprises six emotions - happiness, anger, sadness, disgust, fear, and surprise. These discrete emotions evolved due to their adaptive value for challenges during human phylogeny. Each of the basic emotions is characterized by different situational antecedents, discrete psychological mechanisms and neural signatures as well as distinct behavioral signals (such as facial expressions). While most EEG-ER datasets use a continuous ground truth, some also use categorical labels~\cite{seeda, seedb, seed-iv}.

\subsection{Evaluation Metrics}
\label{sec:Evaluation Metrics}

The most frequently used metrics in EEG-ER are accuracy, the F1 score, and their respecitve standard deviation (std). %
Accuracy is the simplest metric, representing the proportion of correct predictions made by the model over the total number of predictions\cite{grandini2020metricsmulticlassclassificationoverview}. While straightforward, accuracy can be misleading in the context of imbalanced datasets where one class significantly outnumbers others, as a model could achieve high accuracy by simply predicting the majority class for all instances. This is a substantial problem for emotion recognition tasks as most of the datasets are imbalanced in this field.

The F1 score is the harmonic mean of precision and recall \cite{grandini2020metricsmulticlassclassificationoverview}, thus providing a balance between the model's ability to correctly label positive instances (precision) and its capability to find all positive instances (recall). The F1 score reaches its best value at 1 (perfect precision and recall) and its worst at 0. It can be either weighted or unweighted: the unweighted F1 score treats all classes equally, averaging the scores of each class, while the weighted version considers classes proportionally in regards to their presence in the dataset. The unweighted F1 is more informative for imbalanced datasets, as it gives insight into the model's performance across all classes, not just the dominant one.

The standard deviation (std) is used as a measure of the variability or dispersion of the model's performance either across folds in cross-validation~\cite{10.3389/fnins.2021.626277} or across epochs~\cite{RN1019}. A low standard deviation indicates that the model's performance metrics are clustered closely around the mean, suggesting reliability and stability in the model's predictions. Conversely, a high standard deviation suggests a wide range of performance outcomes, indicating potential overfitting to specific data samples or instability in the model's learning process.

\section{Literature Review}

\begin{figure}
    \centering\includegraphics[width=0.5\textwidth]{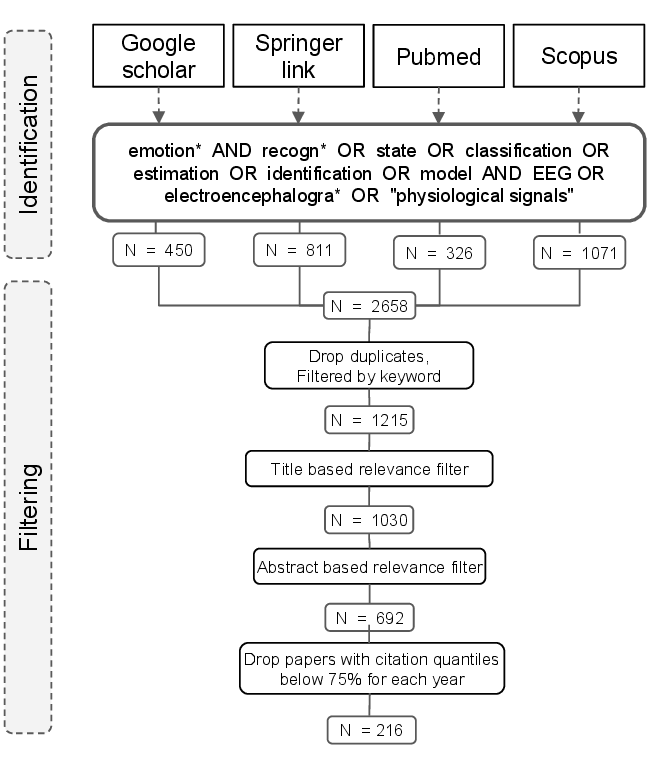}
    \caption{Flow diagram illustrating the inclusion steps of publications into the literature review.}
    \label{fig:flow}
\end{figure}

\begin{figure}
  \centering
  \includegraphics[width=0.5\textwidth,height=\textheight,keepaspectratio]{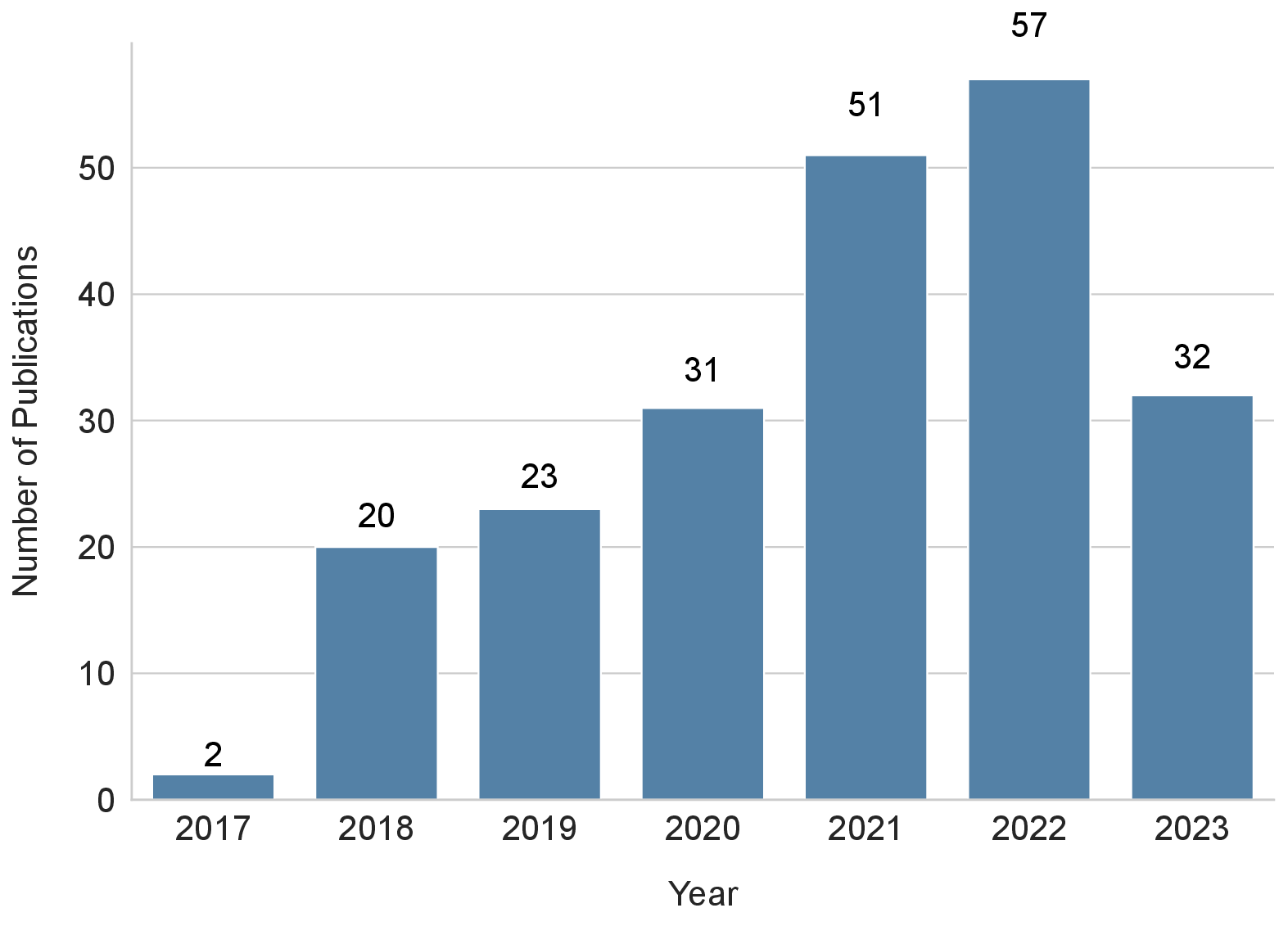}
  \caption{Number of publications per year included in the literature review. In total, 217 publications were included.
  }
  \label{fig:publication_nums}
\end{figure}

We conducted a systematic literature search across four popular databases: Pubmed, Scopus, Springer Link, and Google Scholar (see \autoref{fig:flow} for an overview of the procedure). 
The initial search yielded 2658 papers, which were reduced to 1215 after duplicate removal. 
We further refined the search focusing on those articles that made use of at least one of the most common EEG emotion recognition datasets, namely AMIGOS~\cite{amigos}, DEAP~\cite{deap}, DREAMER~\cite{dreamer}, MAHNOB-HCI~\cite{mahnob}, SEED~\cite{seeda, seedb}, and SEED IV~\cite{seed-iv}.
Finally, we filtered based on citations, including the top 25\% of articles from 2018-2022 and all articles from 2023. Through this strategy, we were able to analyze the 216 most relevant publications in depth (see \autoref{fig:publication_nums}).

 \subsection{Datasets}
 \label{sec:datasets}

\begin{figure}[ht]
  \centering
\includegraphics[width=0.5\textwidth,height=\textheight,keepaspectratio]{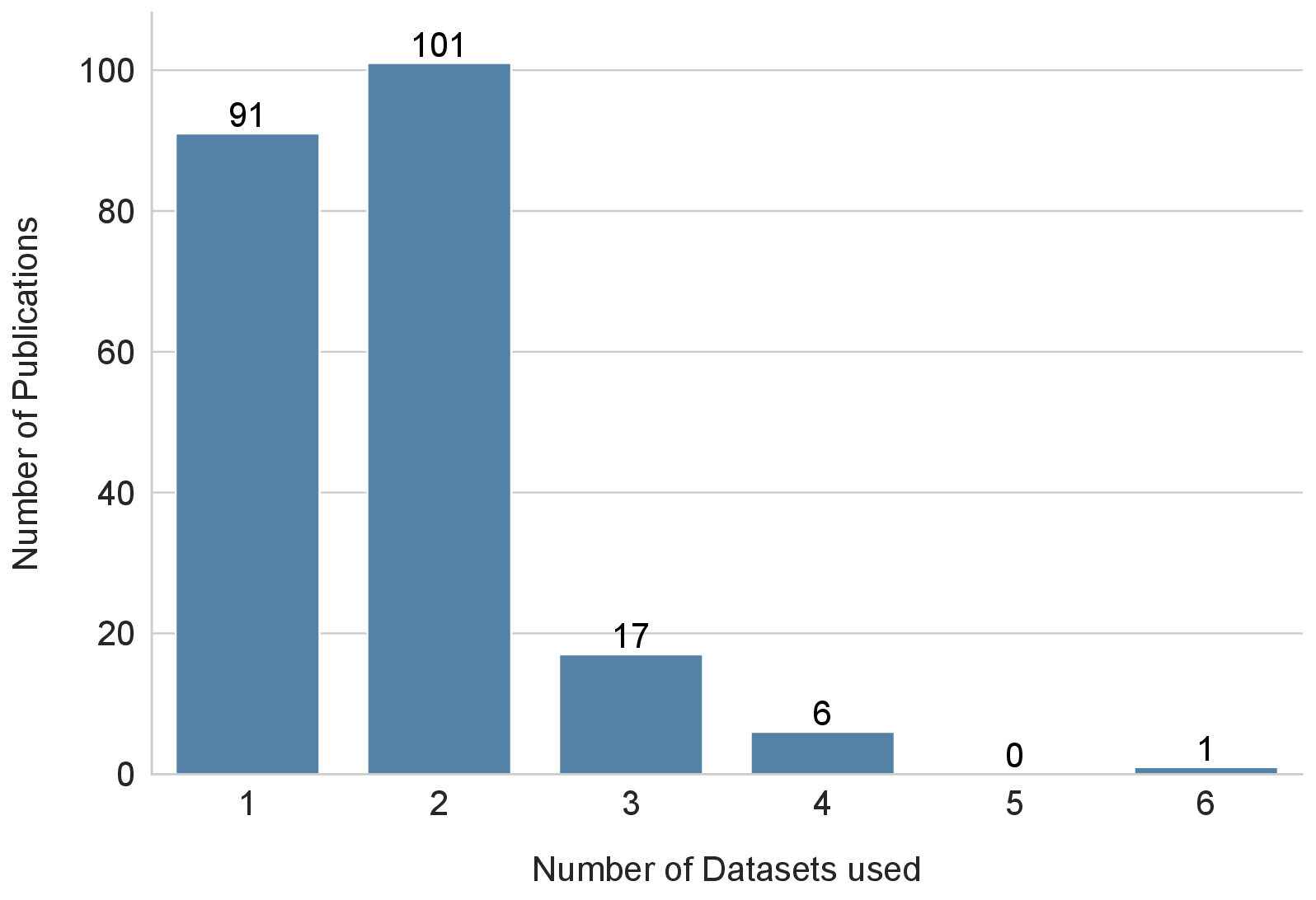}
  \caption{Number of publications using a specific number of different datasets. As can be seen, only a minority of studies used three or more datasets. 
  }
  \label{fig:used_datasets}
\end{figure}

As shown in \autoref{fig:used_datasets}, among the reviewed papers, 42.4\% employ only a single dataset, while 46.5\% utilize two datasets. Only 7.8\% of all papers use three datasets, only 2.7\% integrate four datasets, and only 0.5\% make use of six datasets. Thus, nearly 90\%  of the research is assessed on only one or two datasets, preventing a robust evaluation of the methods.
EEG datasets are characterized by their limited size, different self-assessment scales, and different emotion elicitation strategies as shown in \autoref{tab:dataset_summary} and described herafter.
A research paper may therefore yield good results on one dataset but exhibit notably poorer performance on a different dataset. 
As an example, a proposed method based on bi-hemispheric discrepancy achieved an accuracy of 93.12\% on SEED and  74.35\% on SEED-IV \cite{9105104}.

\begin{table*}[ht]
  \caption{Summary of EEG datasets characteristics }
  \label{tab:dataset_summary}
  \centering
  \begin{tabular}{@{}lllllllll@{}}
    \toprule
    Dataset & Electrode Type & Device Num Channels & Sampling Rate & N Subjects & N Trials & Trial Length & Ground Truth \\
    \midrule
    AMIGOS & Dry & 14 & 128 & 40 & 16 & Approx. 2 min & dimensional\\
    DEAP & Wet & 32 & 512 & 32 & 40 & 1 min & dimensional\\
    DREAMER & Dry & 14 & 128 & 23 & 18 & Approx. 3 min & dimensional\\
    MAHNOB-HCI & Wet & 32 & 256 & 27 & 20 & Approx. 1-2 min & both\\
    SEED & Wet & 62 & 200 & 15 & 15 & Approx. 4 min & categorical\\
    SEED-IV & Wet & 62 & 200 & 15 & 24 & Approx. 2 min & categorical\\
    \bottomrule
  \end{tabular}
\end{table*}

\begin{figure}[ht] 
  \centering
  \includegraphics[width=\columnwidth]{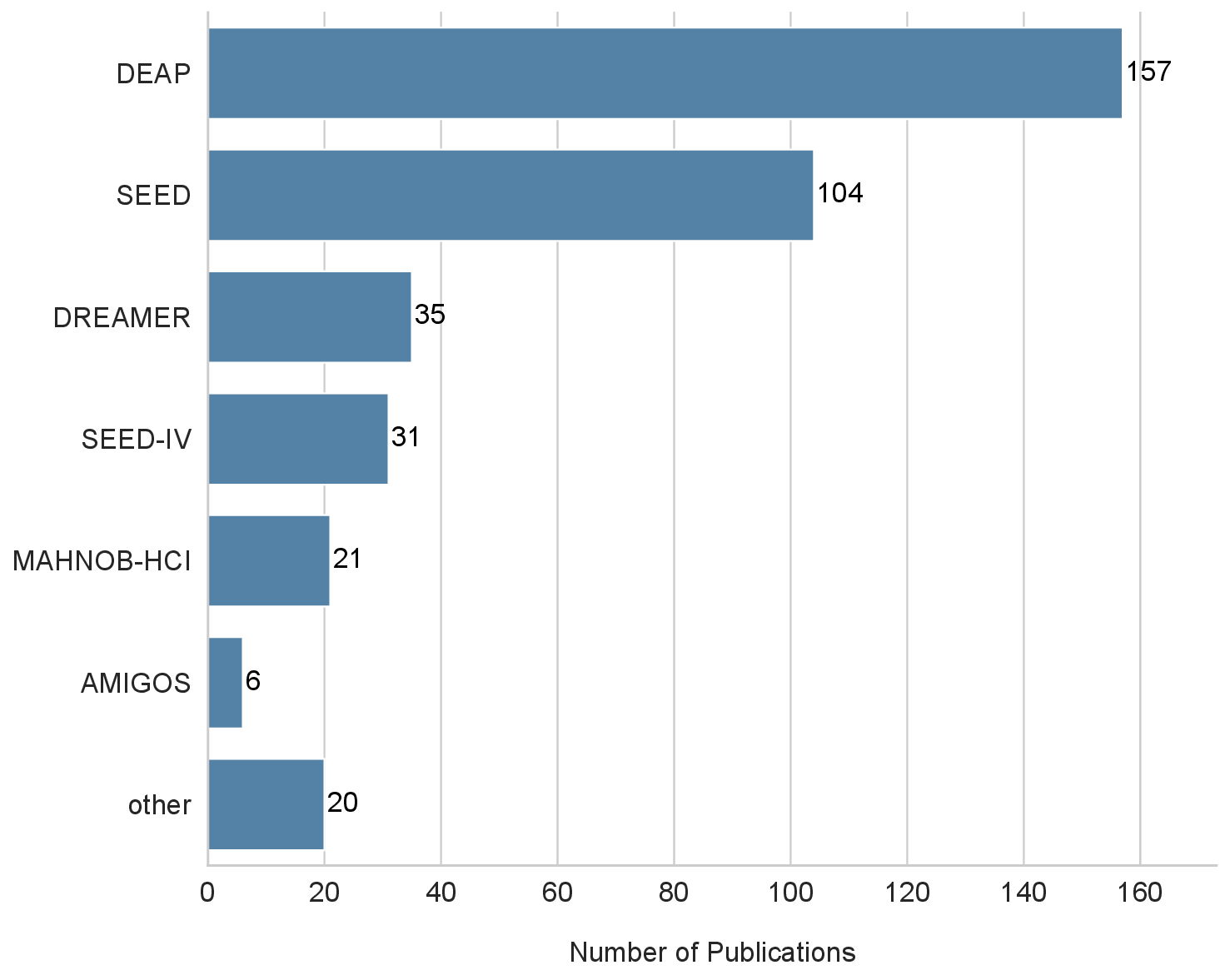} %
  \caption{Number of studies using specific datasets. 
  }
  \label{fig:publication_dataset_nums}
\end{figure}

In the following, we will describe the key characteristics of the six most frequently used datasets. \autoref{fig:publication_dataset_nums} depicts the number of publications employing specific datasets.

 \subsubsection{AMIGOS} The AMIGOS dataset \cite{amigos} is a multimodal dataset which contains EEG data, video recordings of the participants' faces, and peripheral physiological signals. %
 It comprises two experiments. 
    In the first experiment, 40 participants watched 16 emotional video clips on their own.
    The length of the clips varies between 1 and 2 minutes. 
    In the second experiment, 37 of the participants of the previous experiment watched a set of four long affective video extracts from movies. 
    Of these 37 participants, 17 participants watched the videos on their own, whereas 20 participants watched the clips jointly in five groups of four people each. 
    Every participant rated each video in valence, arousal, dominance, familiarity, and liking 
    on a scale ranging from 1 to 9. Furthermore, they selected one basic emotion from the set neutral, happiness, sadness, surprise, fear, anger, and disgust.
 Moreover, three external annotators rated the levels of valence and arousal based on subjects' facial videos for both experiments. 
EEG signals were recorded with 14 channels through dry electrodes (Emotiv EPOC system) at a sampling rate of 128 Hz

 \subsubsection{DEAP} DEAP \cite{deap} is a multimodal dataset which contains EEG, face videos, and peripheral physiological signals.32 volunteers watched a selection of 40 one-minute long excerpts music videos. Every participant rated the videos with respect to valence, arousal, familiarity, liking, and dominance on a scale ranging from 1 to 9. 
 The EEG is recorded using a 32-channel wet electrode system at a 512 Hz sampling rate.
 
 \subsubsection{DREAMER} The DREAMER dataset~\cite{dreamer} is a multimodal database which contains the EEG and peripheral physiological signals of 23 participants 
 as they watched 18 emotional video clips.
  Every participant rated the videos with respect to valence and arousal on a scale ranging from 1 to 5. 
EEG was recorded with a 16-channel dry electrode system (Emotiv EPOC) at a 128 Hz sampling rate.

 \subsubsection{MAHNOB-HCI} MAHNOB-HCI~\cite{mahnob} is a multimodal dataset which contains EEG, face videos, eye gaze, and peripheral physiological signals of 27 participants. 
Every participant rated the videos with respect to valence, arousal, dominance, and predictability on a scale ranging from 1 to 9 and assigned categorical emotional labels. These labels were happiness, amusement, neutral, anger, fear, surprise, anxiety, disgust, and
sadness. 
 EEG was recorded using a 32-channel system with wet electrodes at 256 Hz sampling rate

\subsubsection{SEED} 
The SEED dataset~\cite{seeda, seedb} is a multimodal dataset with EEG recordings of 15 participants and eye-movement recordings of 12 participants. Participants watched 15 emotional movie clips that were chosen based on their ability to elicit positive, negative, and neutral emotional states. Each subject participated in the recordings twice with an interval of at least one week between recordings. After each film clip, participants reported what they felt in response to the clip, whether they had watched the clip before, and whether they understood the clip. Importantly, however, the ground truth labels were not based on these participant reports but consisted of the categorical labels that the researchers assigned to the individual clips (positive, negative, neutral).
EEG was recorded using a 62-channel system with wet electrodes (ESI NeuroScan System). Originally, EEG data were recorded at a 1000 Hz sampling rate, but the data that is available to the public is already downsampled to 200 Hz. 

\subsubsection{SEED-IV} 
SEED-IV~\cite{seed-iv} is a multimodal dataset with EEG and eye-movement recordings of 15 participants.
Each participant underwent three sessions on different days, watching 72 video clips in total (24 per session) while their EEG signals and eye movements were recorded.
These clips were divided equally among four emotion categories: happy, sad, fearful, and neutral, with 18 clips from each category. After each film clip, participants rated the elicited emotions using the PANAS scales~\cite{RN1020}. Importantly, however, the ground truth was not based on these participant reports, but on the discrete labels assigned to each of the film clips (happy, sad, fearful, and neutral). 
The EEG was recorded using a 62-channel system with wet electrodes (ESI NeuroScan System). Originally, EEG data was recorded at a 1000 Hz sampling rate, but the data available to the public is downsampled to 200 Hz - similar to the SEED dataset. %

\begin{figure}[ht]
  \centering
  \includegraphics[width=\linewidth, keepaspectratio]{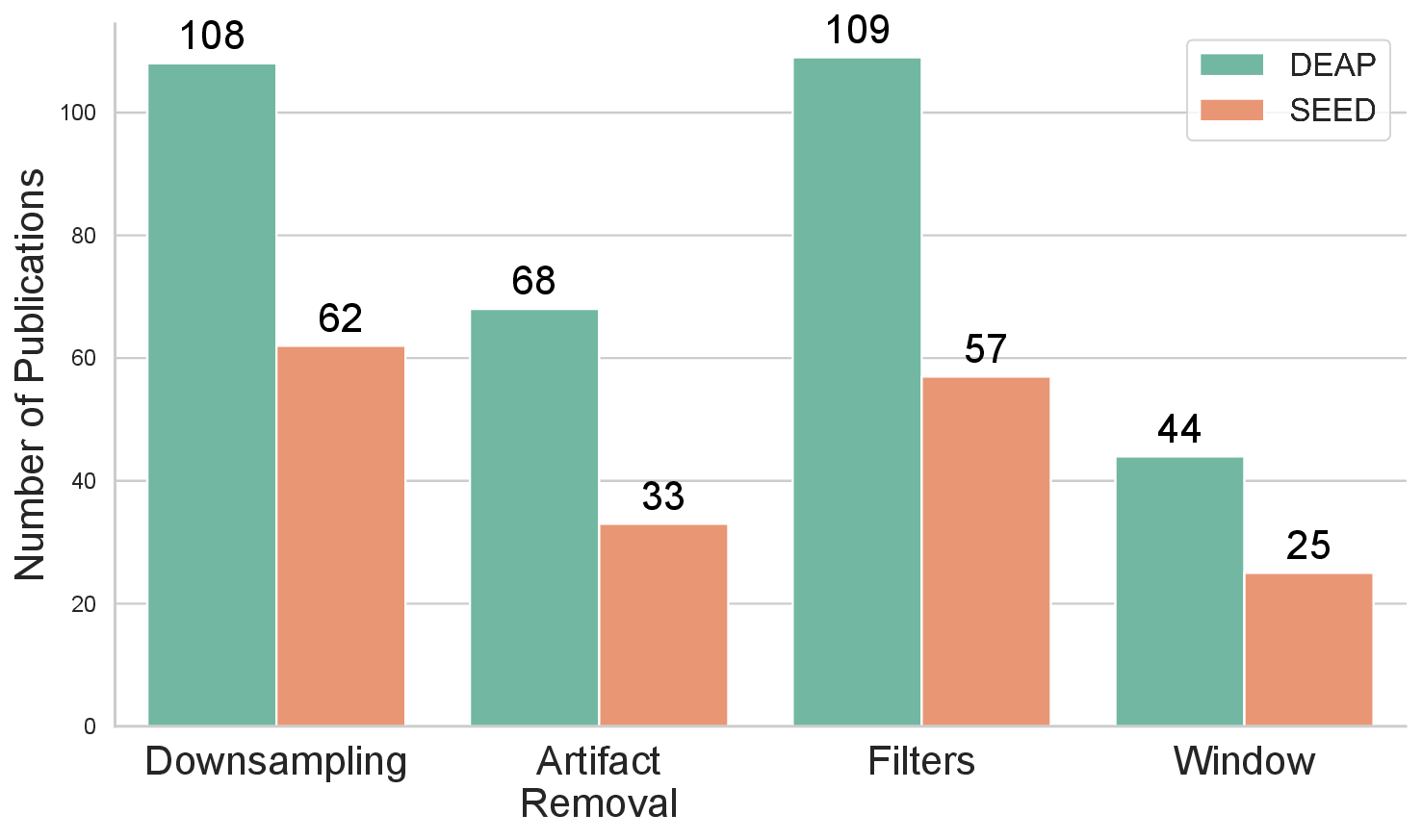}
  \caption{Number of experiments reporting specific pre-procssing steps applied to the DEAP and SEED datasets.
  }
  \label{fig:pre_processing_stats}
\end{figure}

\subsection{Data Pre-processing}

The applied pre-processing techniques, when fully reported, vary substantially between different works (see \autoref{fig:pre_processing_stats}). For example, the data are frequently downsampled to 128 Hz to reduce volume and to equate the sampling rate between different datasets~\cite{2021104757, 7938737, 8489331, 8634938, s19092212, 9204431}. However, some studies downsampled the data to different frequencies such as 256 Hz~\cite{8517037} or did not apply downsampling at all~\cite{Salama2018, 10.1007/978-3-319-70093-9_86}. Subsequently, frequency filters are applied. Frequently, a band-pass filter with a high-pass threshold of 4 Hz and a low-pass threshold of 45 Hz is used~\cite{2021104757, 8489331, 8634938, s19092212, OzdemirDegirmenciIzciAkan+2021+43+57}. However, some studies use high-pass thresholds of 0.5 Hz~\cite{7938737, Bagherzadeh2022} or 1 Hz~\cite{Salama2018} and low-pass thresholds of 64 Hz~\cite{NAKISA2018143}, 70 Hz~\cite{7938737} or no low-pass filter at all~\cite{Salama2018}. Many (but not all) reviewed studies additionally report removal of artifacts such as EOG, EMG, and ECG~\cite{8882370, 7883875, 9321519, 10496191}.  Finally the signal is segmented with a window size that can vary between 1s~\cite{7938737, 8489331} to 15s~\cite{8517037, OzdemirDegirmenciIzciAkan+2021+43+57}. Some studies additionally normalize the data~\cite{8517037, Salama2018, 10.1007/978-3-319-70093-9_86}.

\begin{figure}[ht]
\centering
\includegraphics[width=0.5\textwidth,keepaspectratio]{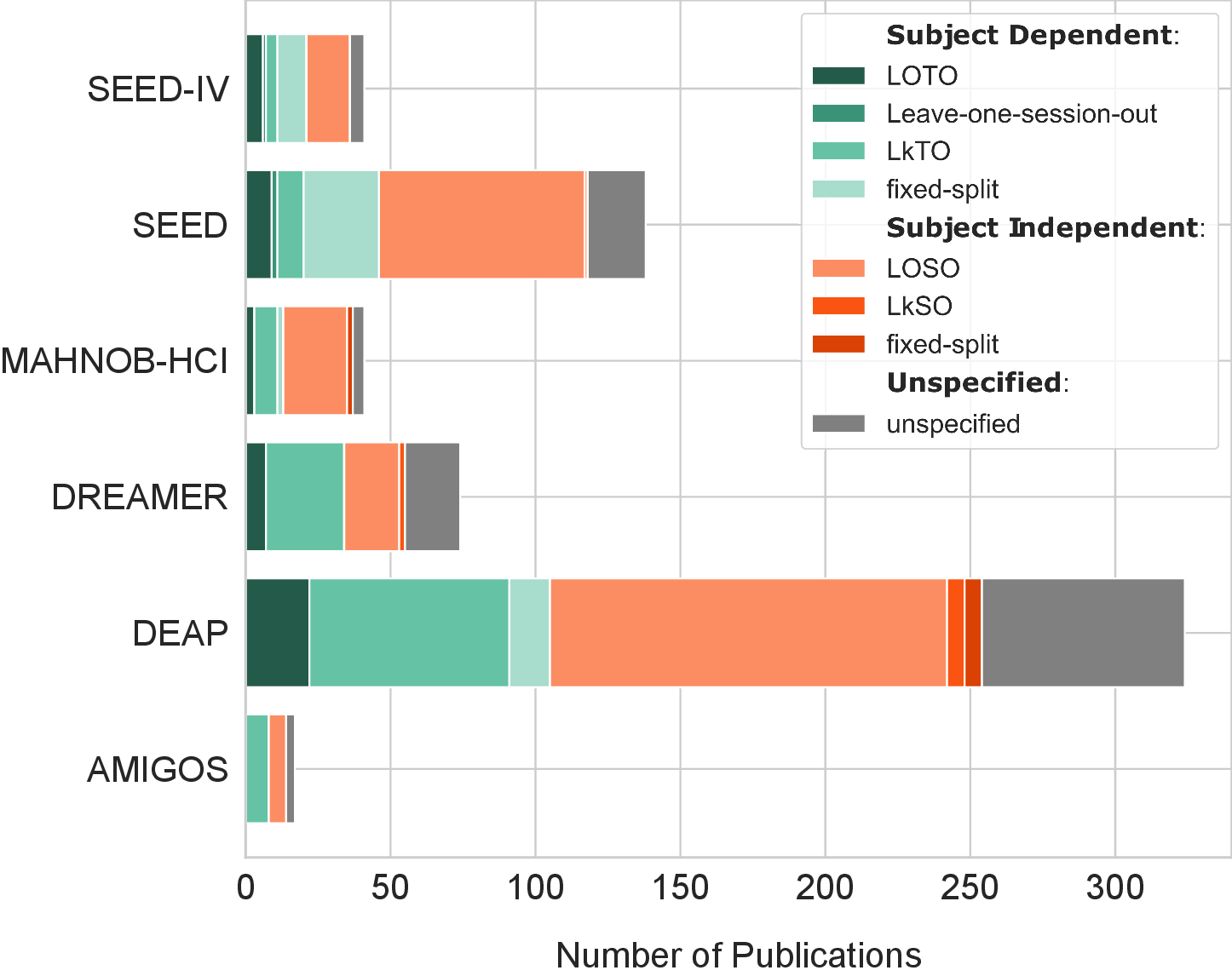}
\caption{Number of publications using specific splitting types, separately for different datasets employed}
\label{fig:data_split_types_across_datasets}
\end{figure}

Taken together, %
the heterogeneity of pre-processing piplines applied to the same datasets poses challenges to comparability. 
If two novel ML approaches are evaluated using different pre-processing steps, it is difficult to ascertain whether any difference between these approaches is attributable to factors such as network architecture, or rather to the pre-processing differences.

 \subsection{Data Splitting}
As illustrated in \autoref{fig:data_split_types_across_datasets}, there is large variability in train/test splitting approaches . Specifically, researchers adopted variants of subject-dependent splitting in 35.7\% of all reviewed experiments and variants of subject-independent splitting in 45.2\%. In 19.1\% of all experiments, not enough information was reported in order to precisely determine the splitting approach. The distribution of splitting approaches varies substantially across datasets.
Thus, approaches using the DEAP dataset more frequently make use of subject-independent splitting compared to approaches on DREAMER. Importantly, even within a given approach and for a given dataset, a variety of different splitting approaches is used.
For subject-independent approaches, while LOSO is the dominant splitting type~\cite{9154557, app122111255, ZHOU2023126262}, some studies use LkSO~\cite{9314183, stBHOSALE2022103289, 10035965}, whereas others use a fixed split~\cite{PANDEY20221730, s19214736, 9024211}. For example, one study held back 2 subjects of the DEAP dataset while the remaining 30 subjects were used for training~\cite{PANDEY20221730}. Many studies applying a subject-dependent approach use LOTO~\cite{8320798, 9154557, 8367876} or LkTO
~\cite{MAHESHWARI2021104428, 8489331, s19092212} cross validation as splitting type while others use a fixed split ~\cite{GUO2022127700, zhong2020eeg, 10496191}. Because the SEED and SEED-IV datasets contain multiple sessions recorded at different times for each subject, some studies working with those datasets use leave-one-session-out cross validation where one session is retained for testing while the remaining sessions are used for training for each iteration ~\cite{9154557, GUO2023104998}.

An attempt to improve generalizability in EEG-ER has been proposed by exploiting domain adaptation techniques~\cite{GUO2023104998, 10.1145/3474085.3475697, LI2023111011, Zhao_Yan_Lu_2021, 8882370}. For example, data are split using one-source-to-one (denoted as O → O) \parencite{Generator}, where one subject becomes the source domain and another becomes the target domain. The (O → O) splitting approach is dynamic, as it involves a systematic variation in the selection of subjects for training and testing. This method entails choosing one subject's data as the training set (source domain) and another subject's data as the testing set (target domain). By iterating this process across different pairs of subjects, the model is tested for its ability to generalize from one individual to another. %

\begin{figure*}
  \centering
  \includegraphics[width=0.8\textwidth]{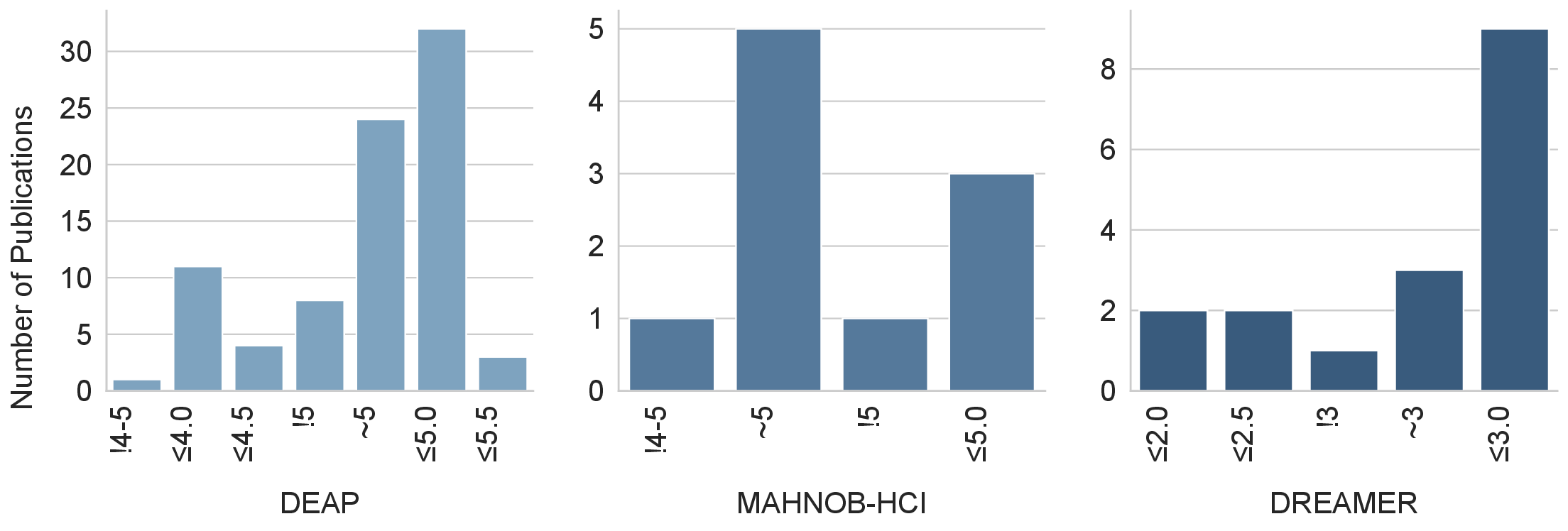}
  \caption{Number of publications employing different threshold values for binarization of valence and arousal ratings into two classes, separately for three different datasets. The tilde symbol indicates that a split is conducted at a specific value without indication whether the value itself falls into the low or the high valence/arousal class. The exclamation mark indicates that samples with a specific value or falling within a specific range were neither included in the low nor in the high class (either completely excluded or included in a neutral class). }
  \label{fig:ground_truth_splitting}
\end{figure*}

\begin{figure}[ht] 
  \centering
  \includegraphics[width=0.5\textwidth,keepaspectratio]{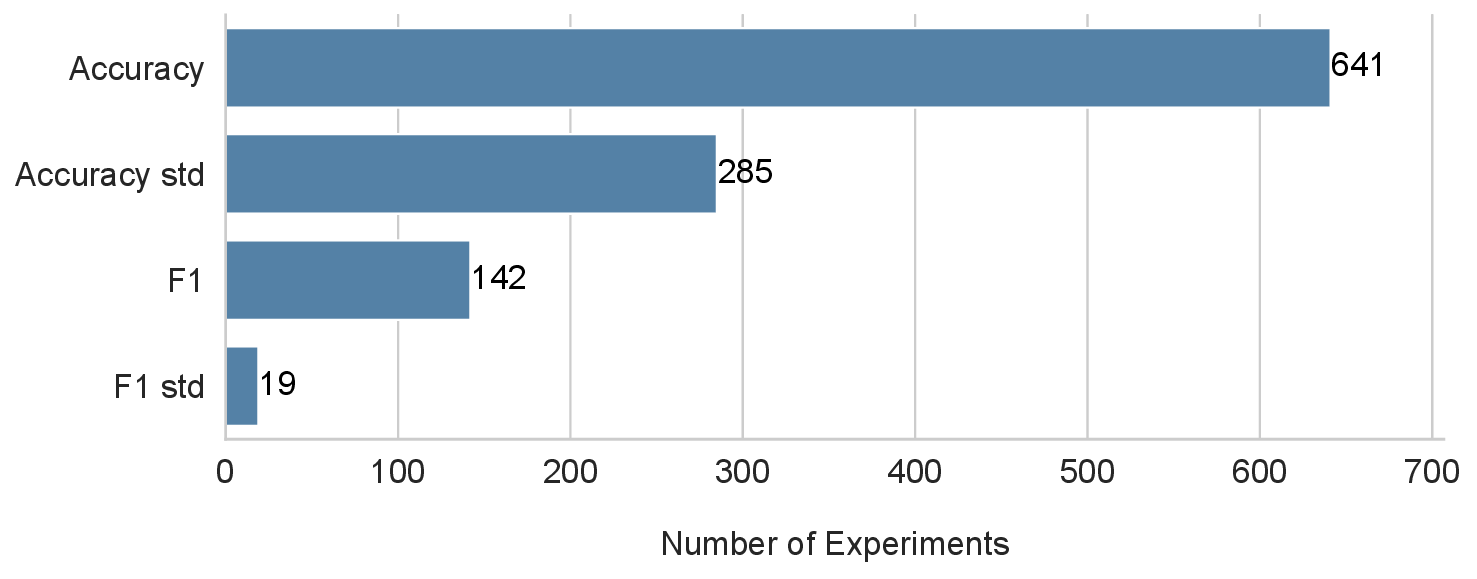} %
  \caption{Number of experiments reporting specific evaluation metrics.
  }
  \label{fig:metrics_stats}
\end{figure}

 \subsection{Ground Truth}
Most of the analyzed studies---specifically those that use the AMIGOS, DEAP, DREAMER, MAHNOB-HCI datasets---aim to classify valence and arousal levels. Importantly, these experiments do generally not predict continuous valence and arousal values but define discrete classes. For example, while most studies define two two-class problems---a binary valence classification and a binary arousal classification---some studies define one four-class problem with the classes low-arousal/low-valence, low-arousal/high-valence, high-arousal/low-valence, high-arousa/high-valence~\cite{wang2022transformers, s21041262, 10035420}, and others define a five-class problem with an additional neutral category~\cite{BAGHERZADEH2022103544}. Critically, even within the same problem definition, different studies use different thresholds to discretize the specific classes.
\autoref{fig:ground_truth_splitting} illustrates the number of studies employing specific threshold values for valence and arousal for different datasets. For example, for the DEAP dataset, more than 30 publications employed a threshold value of less than or equal to 5, while more than 10 publications employed a threshold value of less than or equal to 4. In total, 7 different threshold definitions are employed for the DEAP dataset in the analyzed publications.

This variety of thresholds used in the literature limits the comparability of different approaches because it affects the class distribution of the data. In the DEAP dataset, for example, when binarizing valence and arousal with a threshold of less than or equal to 5, the class ratio of high versus low valence is 56.5\%/43.5\%, respectively, and the class ratio of high vs. low arousal is 58.9\%/41.1\%. However, when the threshold is set to less than or equal to 4, the high/low valence ratio becomes 72.2\%/27.8\%, and the high/low arousal ratio becomes 71.25\%/28.75\%~\cite{deap}. Comparing two approaches using these different threshold values is difficult because trivial baseline performance differs between these two approaches. More specifically, a naive classifier that simply predicts the majority class for all samples would reach a valence-classification accuracy of only 56.5\% in the first case but of 72.2\% in the second case.

\subsection{Metrics}
\autoref{fig:metrics_stats} illustrates the number of experiments reporting specific evaluation metrics. The use of these metrics is not necessarily exclusive or mutually exclusive. As can be seen, 98.8\% (i.e., the vast majority) of all analyzed experiments reports accuracy whereas only 21.9\% of all analyzed experiments report some version of the F1-score. The standard deviation as a measure of model dispersion is reported for accuracy in 43.9\% of all analyzed experiments and for the F1-score in only 2.9\% of all analyzed experiments. It should be noted, however, that only in those experiments that use some form of cross-validation as splitting type the standard deviation can be meaningfully reported. Notably---depending on the applied ground-truth threshold (see \autoref{sec:Ground Truth})---datasets can be severely imbalanced in regard to class distributions. Thus, solely reporting accuracy might not adequately reflect model performance.

\section{Proposed System - EEGain}

\begin{figure*}[ht]
  \centering
  \includegraphics[width=\textwidth,height=\textheight,keepaspectratio]{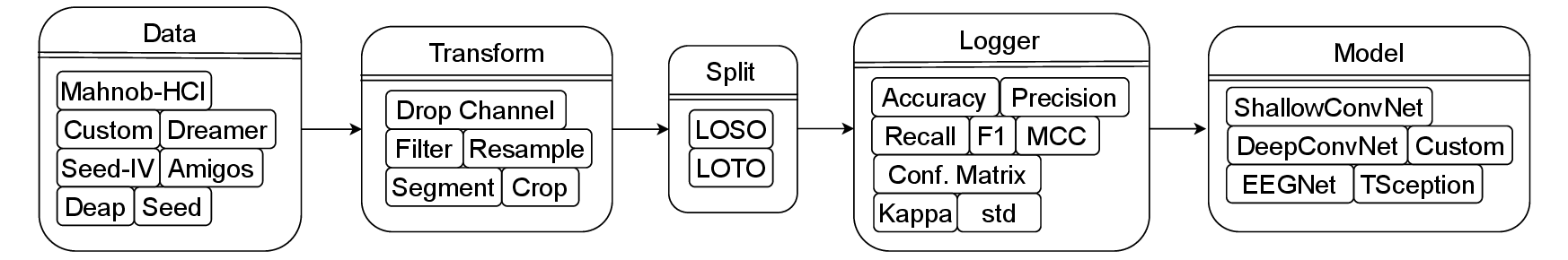}
  \caption{Illustration of EEGAIN's modules.}
  \label{fig:eegain_diagram}
\end{figure*}

In order to overcome the substantial variability in the field of EEG-ER described above, we propose \frameworkname{} to the scientific community.\footnote{Interested researchers can download \frameworkname{} from the following GitHub repository: \url{https://github.com/EmotionLab/EEGain}} \frameworkname{} is the first open-source software that enables a fair and easy comparison among state-of-the-art methods in EEG-ER. 
It handles the entire emotion recognition pipeline, from dataset input to the evaluation of the results. It consists of 5 blocks: Data, Transform, Split, Model, and Logger (see \autoref{fig:eegain_diagram}). The framework can be integrated with Google Colab.

There are two ways to leverage our software. The first way is to write Python code that creates all the instances: a dataset, transformation, split, model class instances, and a training loop function. The second way is to run the experiments from the command line, as reported in our GitHub repository's README file. In this case, the parameters can be directly passed through the command line.

Regarding the training, a helper function (\texttt{\detokenize{def main()}}) is implemented in the framework, with a predefined training loop for LOTO, LOSO, and fixed subject-independent splitting. 
Otherwise, a custom training loop can be created under \texttt{helpers.py}.  

\subsection{Data}
Six widely used datasets can be imported with one line of code. These datasets are: AMIGOS~\cite{amigos}, DEAP~\cite{dreamer}, DREAMER~\cite{dreamer}, MAHNOB-HCI~\cite{mahnob}, SEED~\cite{seeda, seedb}, and SEED-IV~\cite{seed-iv}.  %
A dataset object is created and initialized with the following information: the absolute path where the dataset is stored, the label type (e.g., valence or arousal and categorical emotions), the ground truth threshold, and a list of possible transformations as detailed in the next subsection. %
For AMIGOS, DEAP, DREAMER, and MAHNOB-HCI, binary classification is directly implemented. Data points below or equal to the ground truth threshold will be classified as low (0), while those above will be classified as high (1). 
For SEED and SEED-IV, categorical classification is performed with three (neutral, positive, negative) and four (happiness, sadness, fear, neutral) classes, respectively.

\frameworkname{} can also be used to import new or different datasets. 
All dataset classes are child classes of the \texttt{\detokenize{EEGDataset}} class. When a custom dataset needs to be imported, a child class has to be created and two functions (i.e., $\texttt{\detokenize{__get_subject_ids__}}$ and $\texttt{\detokenize{__get_subject__}}$) need to be defined.

\subsection{Transform}

Transform refers to the pre-processing steps described in the Background section.  In the framework, we have implemented a filtering step consisting of band-pass and notch filters and a windowing step. Additionally, we have added the possibility to crop the signal and delete channels.

\subsection{Split}
Both LOSO and LOTO splitting are available in the framework. LOSO is used for subject-independent studies and LOTO for subject-dependent studies. 
By calling these methods on the dataset object, the dataset is directly split for the training and the test phases.

For the training phase, we have provided a training and validation split mechanism that splits the dataset into training and validation sets in a balanced manner. This is done by sampling data equally from all the classes. The ratio of this training and validation split is set to 0.8 by default, but can be changed by using the appropriate command-line argument when running the framework.

\subsection{Model}
Four main models can be directly run in the framework. These models are DeepConvNet, ShallowConvNet, EEGNet, and TSception. To create a model instance, parameters such as the number of classes, input size, sampling rate, and dropout rate for the training need to be provided. Additionally, a custom model can be created by adding a new class.

The framework also includes two random models, which serve as a trivial baseline against which new models can be tested. The first random model predicts the most occurring class in the training and validation sets. The second random model predicts a random class based on the class distribution in the training and validation sets. We recommend using the former when comparing accuracies and the latter when comparing F1 scores.

\subsection{Logger}
After your model has been trained, logging your results is necessary. For this purpose, a Logger block has been added within the framework to log all the results on TensorBoard. This block consists of two parts: the \texttt{\detokenize{SubjectLogger}}, which can be used to log metrics for each subject/trial, and the \texttt{\detokenize{EmotionLogger}}, which can be used for handling the overall logging functionality by aggregating the results from the \texttt{\detokenize{SubjectLogger}}. The framework provides the capability to log all relevant metrics, including accuracy, F1, weighted F1, Precision, Recall, confusion matrices, Matthew's Correlation Coefficient (MCC), Cohen's Kappa, as well as the standard deviation for all metrics.

\begin{table*}[ht]
  \caption{Pre-processing pipeline for each dataset in the leave-one-subject-out evaluation.}
  \label{tab:preprocess_table}
  \centering
  \begin{tabular}{@{}l>{\raggedright\arraybackslash}p{2.5cm}>{\raggedright\arraybackslash}p{5cm}lll@{}}
    \toprule
    Dataset & Cropping & Channels Dropped & Band-pass Filter & Notch Filter & Ground Truth \\
    \midrule
    MAHNOB-HCI & 30 secs pre and post-baseline &
    "EXG1", "EXG2", "EXG3", "EXG4", "EXG5", "EXG6", "EXG7", "EXG8", "GSR1", "GSR2", "Erg1", "Erg2", "Resp", "Temp", "Status" &
    [0.3Hz, 45Hz] & 50Hz & $\leq4.5$ \\
    DEAP & 3 secs pre-baseline & "EXG1", "EXG2", "EXG3", "EXG4", "GSR1", "Plet", "Resp", "Temp" & - & 50Hz & $\leq4.5$ \\
    AMIGOS & - & "ECG\_Right", "ECG\_Left", "GSR" & - & 50Hz & $\leq4.5$ \\
    DREAMER & - & - & [0.3Hz, 45Hz] & 50Hz & $\leq3$ \\
    SEED & - & - & [0.3Hz, 45Hz] & 50Hz & - \\
    SEED-IV & - & - & [0.3Hz, 45Hz] & 50Hz & - \\
    \bottomrule
  \end{tabular}
\end{table*}

\begin{table}[ht]
  \caption{Comparison of TSception implemented in our \frameworkname{} with the originally reported results \cite{tsception2022} and additional previous works for leave-one-trial-out cross-validation on DEAP.}
  \label{tab:DEAP_loto_results}
  \centering
\begin{tabular}{lccccc}
\toprule
 & \multicolumn{2}{c}{\textbf{Arousal}} & \multicolumn{2}{c}{\textbf{Valence}} \\
\cmidrule(lr){2-3} \cmidrule(lr){4-5}
\textbf{Method}  & \textbf{ACC} & \textbf{F1} & \textbf{ACC} & \textbf{F1} \\
\midrule
SVM~\cite{deap}            & 62.00\% & 58.30\% & 57.60\% & 56.30\% \\
UL~\cite{LIANG2019257}             & 62.34\% & 60.44\% & 56.25\% & 61.25\% \\
CSP~\cite{Appriou2020ModernMA}            & 58.26\% & --      & 57.59\% & --      \\
FBCSP~\cite{Appriou2020ModernMA}          & 59.13\% & --      & 59.19\% & --      \\
FgMDM~\cite{Appriou2020ModernMA}          & 60.04\% & --      & 58.87\% & --      \\
TSC~\cite{Appriou2020ModernMA}            & 60.04\% & --      & 59.47\% & --      \\
FBFgMDM~\cite{Appriou2020ModernMA}        & 60.30\% & --      & 61.01\% & --      \\
FBTSC~\cite{Appriou2020ModernMA}          & 60.60\% & --      & 61.09\% & --      \\
\midrule
TSception \cite{tsception2022} & 63.75\% & 63.35\% & 62.27\% & 65.37\% \\
\midrule
TSception (\frameworkname{}) & 60.67\% & 61.40\% & 59.32\% & 62.49\% \\
Trivial Baseline (\frameworkname{}) & 62.73\% & 55.82\% & 50.31\% & 53.81\% \\
\bottomrule
\end{tabular}
\label{tab:loto_results}
\end{table}

\subsection{Prediction Logging}
Additionally, there is a functionality within the framework that allows the user to log the predictions from the test sets. The said functionality is defined within the helper functions for the LOSO and LOTO training loops. This functionality was added to aid the users in performing more complex analyses when dealing with EEG data and the corresponding audio/video stimuli. The predictions are stored in CSV-files and provide the user with all the necessary information to extract data for any particular training epoch, video ID, or subject ID that they want to inspect or manipulate for further analyses.

\section{Experiments}

\begin{table*}[h]
  \caption{Leave-one-subject-out evaluation results on different datasets.}
  \label{tab:LOSO_results}
  \centering
  \begin{tabular}{llllll}
    \toprule
   Dataset & Task & Model & Accuracy & F1 & F1 Weighted\\
    \midrule
    Mahnob & Arousal & TSception & \textbf{0.54 $\pm$ 0.11}  & \textbf{0.49 $\pm$ 0.24} & 0.51 $\pm$ 0.12\\
                               &  & EEGNet & 0.52 $\pm$ 0.11  & 0.43 $\pm$ 0.21 & 0.49 $\pm$ 0.12\\
                               &  & DeepConvNet & \textbf{0.54 $\pm$ 0.13} & 0.44 $\pm$ 0.27 & 0.49 $\pm$ 0.16\\
                               &  & ShallowConvNet & 0.52 $\pm$ 0.11 & 0.44 $\pm$ 0.21 & 0.49 $\pm$ 0.12\\
                               &  & Trivial Baseline & 0.34 $\pm$ 0.11 & 0.48 $\pm$ 0.11 & \textbf{0.52 $\pm$ 0.04}\\ [0.01cm]
                               \cline{2-6}
                               \vspace{-0.2cm}\\
                               & Valence & TSception & 0.53 $\pm$ 0.07  & 0.52 $\pm$ 0.16 & 0.50 $\pm$ 0.08\\
                               &  & EEGNet & 0.56 $\pm$ 0.08  & \textbf{0.58 $\pm$ 0.15} & 0.53 $\pm$ 0.10\\
                               &  & DeepConvNet & 0.56 $\pm$ 0.08 & 0.56 $\pm$ 0.18 & 0.53 $\pm$ 0.11\\
                               &  & ShallowConvNet & \textbf{0.57 $\pm$ 0.08} & 0.57 $\pm$ 0.18 & \textbf{0.54 $\pm$ 0.10}\\
                               &  & Trivial Baseline & 0.55 $\pm$ 0.10 & 0.53 $\pm$ 0.06 & 0.50 $\pm$ 0.03\\
    \midrule
    Deap & Arousal & TSception & 0.54 $\pm$ 0.09  & 0.56 $\pm$ 0.20 & \textbf{0.53 $\pm$ 0.10}\\
                               &  & EEGNet & 0.53 $\pm$ 0.13  & 0.53 $\pm$ 0.21 & 0.50 $\pm$ 0.13\\
                               &  & DeepConvNet & 0.57 $\pm$ 0.14 & 0.52 $\pm$ 0.31 & 0.49 $\pm$ 0.16\\
                               &  & ShallowConvNet & 0.56 $\pm$ 0.13 & 0.57 $\pm$ 0.26 & 0.49 $\pm$ 0.13\\
                               &  & Trivial Baseline & \textbf{0.59 $\pm$ 0.15} & \textbf{0.58 $\pm$ 0.08} & \textbf{0.53 $\pm$ 0.04}\\[0.01cm]
                               \cline{2-6}
                               \vspace{-0.2cm}\\
                               & Valence & TSception & 0.51 $\pm$ 0.08  & 0.55 $\pm$ 0.16 & 0.47 $\pm$ 0.08\\
                               &  & EEGNet & 0.53 $\pm$ 0.10  & 0.56 $\pm$ 0.21 & 0.45 $\pm$ 0.12\\
                               &  & DeepConvNet & 0.51 $\pm$ 0.11 & 0.47 $\pm$ 0.28 & 0.41 $\pm$ 0.13\\
                               &  & ShallowConvNet & 0.53 $\pm$ 0.08 &  \textbf{0.61 $\pm$ 0.18} & 0.45 $\pm$ 0.12\\
                               &  & Trivial Baseline & \textbf{0.57 $\pm$ 0.09} & 0.56 $\pm$ 0.05 & \textbf{0.51 $\pm$ 0.03}\\
    \midrule
    Amigos & Arousal & TSception & 0.58 $\pm$ 0.18  & 0.64 $\pm$ 0.24 & 0.57 $\pm$ 0.20\\
                               &  & EEGNet & 0.60 $\pm$ 0.23  & 0.69 $\pm$ 0.24 & 0.56 $\pm$ 0.25\\
                               &  & DeepConvNet & 0.56 $\pm$ 0.21 & 0.65 $\pm$ 0.24 & 0.55 $\pm$ 0.22\\
                               &  & ShallowConvNet & 0.59 $\pm$ 0.23 & \textbf{0.70 $\pm$ 0.23} & 0.55 $\pm$ 0.25\\
                               &  & Trivial Baseline & \textbf{0.66 $\pm$ 0.26} & 0.62 $\pm$ 0.19 & \textbf{0.59 $\pm$ 0.11}\\[0.01cm]
                               \cline{2-6}
                               \vspace{-0.2cm}\\
                               & Valence & TSception & 0.53 $\pm$ 0.10  & 0.56 $\pm$ 0.18 & 0.51 $\pm$ 0.13\\
                               &  & EEGNet & 0.55 $\pm$ 0.11  & 0.59 $\pm$ 0.19 & 0.50 $\pm$ 0.14\\
                               &  & DeepConvNet & 0.55 $\pm$ 0.11 & 0.56 $\pm$ 0.18 & \textbf{0.52 $\pm$ 0.13}\\
                               &  & ShallowConvNet & 0.55 $\pm$ 0.13 & \textbf{0.60 $\pm$ 0.20} & 0.51 $\pm$ 0.15\\
                               &  & Trivial Baseline & \textbf{0.56 $\pm$ 0.14} & 0.55 $\pm$ 0.07 & 0.51 $\pm$ 0.05\\
    \midrule
    Dreamer & Arousal & TSception & 0.47 $\pm$ 0.07  & 0.43 $\pm$ 0.14 & 0.46 $\pm$ 0.08\\
                               &  & EEGNet & 0.46 $\pm$ 0.09  & \textbf{0.47 $\pm$ 0.19} & 0.42 $\pm$ 0.13\\
                               &  & DeepConvNet & 0.48 $\pm$ 0.11 & 0.41 $\pm$ 0.18 & 0.45 $\pm$ 0.12\\
                               &  & ShallowConvNet & 0.48 $\pm$ 0.08 & 0.41 $\pm$ 0.18 & 0.45 $\pm$ 0.10 \\
                               &  & Trivial Baseline &  \textbf{0.52 $\pm$ 0.15}  &  0.46 $\pm$ 0.07  &  \textbf{0.51 $\pm$  0.02}\\[0.01cm]
                               \cline{2-6}
                               \vspace{-0.2cm}\\
                               & Valence & TSception & \textbf{0.60 $\pm$ 0.06}  & \textbf{0.42 $\pm$ 0.15} & \textbf{0.57 $\pm$ 0.07}\\
                               &  & EEGNet & \textbf{0.60 $\pm$ 0.08}  & 0.26 $\pm$ 0.20 & 0.53 $\pm$ 0.11\\
                               &  & DeepConvNet & \textbf{0.60 $\pm$ 0.09} & 0.38 $\pm$ 0.20 & 0.56 $\pm$ 0.10\\
                               &  & ShallowConvNet &  \textbf{0.60 $\pm$  0.08} &  0.35 $\pm$ 0.19  & 0.56 $\pm$ 0.10 \\
                               &  & Trivial Baseline & 0.59 $\pm$ 0.09 & 0.40 $\pm$ 0.04 & 0.51 $\pm$ 0.03\\
    \midrule
    SEED & Categorical & TSception & 0.48 $\pm$ 0.07  & 0.46 $\pm$ 0.08 & 0.46 $\pm$ 0.08 \\
                               &  & EEGNet & 0.46 $\pm$ 0.06  & 0.44 $\pm$ 0.06 & 0.44 $\pm$ 0.06 \\
                               &  & DeepConvNet & \textbf{0.55 $\pm$ 0.08} & \textbf{0.52 $\pm$ 0.10} & \textbf{0.52 $\pm$ 0.10} \\
                               &  & ShallowConvNet & 0.49 $\pm$ 0.06 & 0.47 $\pm$ 0.07 & 0.47 $\pm$ 0.07\\
                               &  & Trivial Baseline & 0.34 $\pm$ 0.00 & 0.34 $\pm$ 0.02 & 0.34 $\pm$ 0.02 \\
    \midrule
    SEED-IV & Categorical & TSception & 0.40 $\pm$ 0.08  & 0.33 $\pm$ 0.10 & 0.35 $\pm$ 0.10\\
                               &  & EEGNet & 0.32 $\pm$ 0.03  & 0.20 $\pm$ 0.06 & 0.22 $\pm$ 0.05 \\
                               &  & DeepConvNet & \textbf{0.45 $\pm$ 0.09} & \textbf{0.42 $\pm$ 0.11} & \textbf{0.42 $\pm$ 0.11} \\
                               &  & ShallowConvNet & 0.37 $\pm$ 0.06 & 0.32 $\pm$ 0.06 & 0.33 $\pm$ 0.07 \\
                               &  & Trivial Baseline & 0.31 $\pm$ 0.00 & 0.24 $\pm$ 0.03 & 0.24 $\pm$ 0.03 \\
    \bottomrule
  \end{tabular}
\end{table*}

Using \frameworkname{}, we perform two different experimental evaluations.
First, we validate our framework by evaluating our implementation of TSception with the numbers reported by the original authors~\cite{tsception2022}.
Second, we present experimental evaluations of four popular EEG emotion recognition approaches %
in the challenging leave-one-subject-out (LOSO) scenario on six different EEG emotion recognition datasets.
To the best of our knowledge, these are the to date most comprehensive experiments in the LOSO scenario in terms of the number of included datasets.
In this way, we shed light on the ability of popular EEG emotion recognition approaches to generalise across different users, and also highlight the utility of \frameworkname{} for conducting comprehensive experiments.

\subsection{Metrics and Baselines}
We report all results in three evaluation metrics: Accuracy, F1 and F1 weighted.
The F1 score always relates to the positive class.
For example, in a binary classification scenario of \textit{high} vs. \textit{low} arousal, high arousal is the positive class, and low arousal the negative class.
In the case of multi-class classification, 
We compute macro-averaged F1 scores.
For weighted F1 scores, we first compute class-specific F1 scores and then average each class-specific F1 score weighted by the support of the class.
To investigate whether EEG emotion recognition methods can outperform trivial classifiers, we implement the following baselines.
As a trivial baseline for the Accuracy metric, we always predict the most likely class on the training and validation sets.
On the other hand, for F1 and weighted F1, we sample predictions according to the class distributions on the training and validation sets.

\subsection{Implementation Details}

For the validation experiments on DEAP~\cite{deap}, we followed the experimental protocol described in~\cite{tsception2022} as closely as possible.
The 3-second pre-trial baseline was removed for each trial. Then the data was down-sampled from 512Hz to 128Hz. A band-pass filter from 4.0-45 Hz was applied to the original EEG to remove the low and high-frequency noise. A ground truth threshold of 5 was chosen to binarize the labels. The ratio coefficients of the T kernel length are [0.5, 0.25, 0.125] for DEAP. We also dropped the following channels - "EXG1", "EXG2", "EXG3", "EXG4", "GSR1", "Plet", "Resp", "Temp", "Oz", "Pz", "Fz", "Cz" to match the channels used in \cite{tsception2022}.
For model training, the maximum training epoch is 500. The batch size on the DEAP dataset is set to 64. Adam optimizer is utilized to optimize the training process with the initial learning rate being $1e-3$. Cross-entropy loss is selected as the loss function to guide the training process. 

For our comprehensive LOSO experiments, we aimed for a consistent approach across all datasets to test the ability for cross-subject generalization in a unified way.
We performed pre-processing transformations using the \frameworkname{} framework as shown in \autoref{tab:preprocess_table}. 
We binarized the MAHNOB-HCI, DEAP, and AMIGOS labels by using $\leq4.5$ as a threshold (original scale form 1 to 9). For DREAMER, we used $\leq3.0$ (original scale 1 to 5). The SEED and SEED-IV datasets have categorical labels, so no binarization was needed. All datasets were resampled using a sampling rate of 128Hz. Segments of the signal are created using a window size of 4 with an overlap of 0.
All experiments were run for 200 epochs, with a batch size of 32. The learning rate used was 0.001 with no weight decay. For the Cross-entropy loss function, a label smoothing of 0.01 was used as we found that it slightly increased Accuracies ($\approx1\%$) for some models. 
For training the different methods, the data was split by subject into 80\% training and 20\% validation sets.
In each fold of the LOSO loop, we select the model with the best accuracy on the validation set for evaluation on the test subject.

For further implementation details, please refer to the open-access GitHub repository for \frameworkname{}.

\subsection{Validation Results}
\autoref{tab:loto_results} shows the results of our validation experiment for leave-one-trial-out cross-validation on DEAP.
We compared our implementation of TScpetion to the numbers reported in the original paper~\cite{tsception2022}.
For Arousal, we reached an F1 score of 61.40\%, while the authors of \cite{tsception2022} reported 63.35\%.
For Valence, we reached an F1 score of 62.49\%, while 65.37\% were originally reported.
The differences of 1.95\% and 2.88\% respectively are below the average difference of 2.95\% in replication experiments found by~\cite{liu2024libeercomprehensivebenchmarkalgorithm}.
As such, these results successfully validate  \frameworkname{}.

\subsection{LOSO Results}

We present the results of our evaluation in \autoref{tab:LOSO_results}.
Cross-subject prediction in EEG emotion recognition is generally challenging for all investigated architectures.
For many datasets, accuracies and F1 scores do not improve over trivial baselines.
In the case of the DREAMER dataset, none of the evaluated approaches is able to outperform the trivial baselines for arousal prediction.
For Valence prediction, we observe marginal improvements in Accuracy (0.60 vs. 0.59) for EEGNet, DeepConvNet, and ShallowConvNet, as well as stronger improvements in weighted F1 scores (0.57 vs. 0.52) for ShallowConvNet.
On MAHNOB, TSception and DeepConvNet achieve consistent but small improvements in Arousal, while ShallowConvNet achieves slightly more substantial improvements for Valence predictions.
On DEAP and AMIGOS, we observe some but never consistent improvements in F1 score and F1 weighted.
Taken together, the improvements over trivial baselines for datasets with Arousal and Valence estimation tasks are small, if not non-existent.

A different picture emerges for the categorical prediction tasks on SEED and SEED-IV.
Here, we observe substantial improvements over trivial baselines.
On both datasets, DeepConvNet achieves the best results throughout all metrics.
For SEED, DeepConvNet reaches 0.52 weighted F1, clearly outperforming the trivial baseline at 0.34 weighted F1.
For SEED-IV, we observe similar strong gains of 0.42 weighted F1 for DeepConvNet versus 0.24 weighted F1 for the trivial baseline.
Apart from DeepConvNet, other approaches also reach clear improvements over trivial baselines on SEED and SEED-IV.
The only exception is EEGNet on SEED-IV, which is not able to outperform the trivial baselines.

\label{rel_work}

\section{Discussion}
Taken together, the contributions of our paper are threefold. First, our literature review of 216 papers shows that there is substantial variability in the evaluation of EEG-based emotion-recognition models regarding key aspects such as dataset selection, data pre-processing, data splitting, ground-truth definition, and evaluation metrics. While our systematic review only includes articles published up to June 2023, these inconsistencies also occur in more recent publications. For example, the problem of inconsistent dataset selection persists also in studies published in 2025: One study used DEAP and SEED~\cite{ZANG2025107622}, another used DEAP and DREAMER~\cite{Yin17032025}, and one used DEAP, DREAMER, and SEED-IV~\cite{10.3389/fnbot.2024.1481746}. Furthermore, two articles from 2024 testing subject-independent approaches on the SEED dataset~\cite{seeda, seedb} used different data splitting approaches with one study using a fixed split with one subject in the test set and 14 subjects in the training set~\cite{RN1034} and another study using a fixed 80-20 split~\cite{RN1035}. Moreover, two studies published in 2024 working with the DEAP dataset~\cite{deap} use different ground-truth definitions with one study assigning valence and arousal values equal to or larger than 5 to the high class~\cite{XU2024107927} whereas in another study, only values larger than 5 were assigned to the high class (with values equal to 5 excluded)~\cite{RN1036}. Thus, we assume that inconsistencies regarding evaluation approaches are impeding progress in the field to this day. 

The second contribution of our paper is \frameworkname{}, an open source software framework that offers the capacity to address the inconsistencies in EEG-ER evaluation observed in our review by providing easy-to-use and standardized solutions for the whole training and evaluation pipeline. Recently, two frameworks similar to \frameworkname{} were published, TorchEEG\textsubscript{EMO}~\cite{ZHANG2024123550} and LibEER~\cite{liu2024libeercomprehensivebenchmarkalgorithm}. However, we believe that our framework has some advantages over these existing frameworks: 

While LibEER~\cite{liu2024libeercomprehensivebenchmarkalgorithm} provides a foundational structure for EEG-based emotion recognition research, our proposed framework, \frameworkname{}, introduces several enhancements that address key limitations and improve overall usability and extensibility. \frameworkname{} incorporates robust metric and loss tracking through integrated TensorBoard support, facilitating detailed model performance monitoring. Additionally, it supports structured logging of model predictions, enabling downstream analyses such as error inspection, class-distribution analysis, etc. Our framework also provides trivial baseline models, such as majority class prediction and random sampling based on class distribution, to establish meaningful lower bounds for performance evaluation. These baselines are essential for contextualizing model results and ensuring scientifically rigorous comparisons, particularly in imbalanced EEG datasets. \frameworkname{} includes out-of-the-box support for a broader set of publicly available and widely used EEG datasets, ensuring greater coverage and relevance for benchmarking. This enables researchers to more easily compare model generalization across diverse recording conditions and subject populations. Our framework emphasizes automation and reproducibility through its script-driven design. Users can execute complete experiments via shell scripts that internally handle all necessary components, using parameters defined in centralized configuration files. This removes the need for manual imports or code-level changes during routine experimentation.

TorchEEG\textsubscript{EMO}~\cite{ZHANG2024123550} also lacks any intrinsic training or validation monitoring system like the \frameworkname{} framework has with built-in TensorBoard support. Our framework is also more user-friendly, with minimal to no coding required to run experiments.

As a third contribution of our paper, we systematically conducted subject-independent EEG-ER experiments with four state-of-the-art models on six popular datasets using LOSO cross-validation.  To the best of our knowledge, these are the to date most comprehensive
experiments in the LOSO scenario in terms of the number of included datasets. These results can be used by the scientific community as baselines to compare their own models to. 

Taken together, the present review aims to overcome the variability in evaluation approaches in the field of EEG-ER and to accelerate progress in the field by contributing to the unification of evaluation protocols. To this end, we suggest the following guidelines for researchers working on EEG-ER: First, we advise researchers to evaluate their models on multiple datasets to ascertain that their models generalize well to substantially different data. Obviously, not every dataset is suitable to evaluate every model. For example, a model that is trained to perform binary valence and arousal classifications cannot be properly evaluated on SEED~\cite{seeda, seedb} and-SEED-IV~\cite{seed-iv} because these two datasets are characterized by three-class and four-class problems, respectively. Moreover, researchers might struggle to get access to specific datasets, especially when results need to be obtained fast due to submission deadlines. Thus, it might be too demanding to ask researchers to evaluate their models on all six datasets used in the present review. Importantly, however, once researchers have selected the datasets they would like to use to train and/or evaluate their models, they should adhere to dataset-specific standards regarding the following steps of the evaluation pipeline (i.e., pre-processing, data splitting, ground-truth definition, and evaluation metric selection). 

We do not believe that it would be expedient to advise researchers to use only fixed pre-processing methods for specific data sets, as the scientific contribution of a study might precisely lie in the use of novel pre-processing methods. For example, several studies have extracted specific frequency bands from the EEG signal in order to classify emotions~\cite{s19092212, s21041262, BAGHERZADEH2022103544}. For these studies, it would be inherently problematic to use a pre-defined band-pass filter. Importantly, however, we believe that two standards regarding pre-processing techniques should be met by future studies. First, studies have to meticulously report the pre-processing techniques they applied to the raw signals in order to allow future studies to replicate these steps. Second, when a study proposes a new machine-learning architecture and compares its performance to previously published architectures, researchers should mimic the pre-processing steps employed in those previous publications so any potential increases in classification performance can be unambigously attributed to differences in the architecture and not to confounded differences in the pre-processing techniques applied to the raw EEG signal. 

Regarding the issue of splitting data into a training set and a test set, we believe that both the subject-dependent approach and the subject-independent approach have their merit. More specifically, we argue that subject-dependent and subject-independent EEG-ER should be viewed as separate problems. Thus, it should be left to individual researchers to decide whether they want to contribute new methods, models, and algorithms to the field of subject-dependent or subject-independent EEG-ER. Importantly, however, we do believe that within these two domains, researchers should adhere to unified splitting methods because otherwise the performance of different models from the same domain cannot be meaningfully compared. We decided to implement LOSO cross-validation as a subject-independent approach and LOTO cross-validation as a subject-depdentend approach because these approaches were the most frequently used ones in the analyzed publications of our literature review. Moreover, we believe that these splitting methods are beneficial because they allow to use the maximal amount of data for training. We acknowledge that these splitting methods are computationally expensive because they require to train \textit{N} models (with \textit{N} being the number of subjects/trials), but---given the rapid advances in hardware with regard to computational capabilities---we believe that this is an acceptable downside.

Differences in the ground-truth definition (i.e., in the threshold values used to binarize contiunous valence and arousal ratings) affect class distributions and thus trivial-baseline performance. Consequently, one model can achieve better performance than another model simply by using a different ground-truth defintion. Therefore, we believe it to be of utmost importance for researchers to use unified ground-truth definitions in their studies. We propose to use the default values of \frameworkname{} for ground-truth binarization, that is a low-class threshold of smaller than or equal to 3.0 for the DREAMER dataset~\cite{dreamer} and of smaller than or equal to 4.5 for AMIGOS~\cite{amigos}, DEAP~\cite{deap}, and MAHNOB-HCI~\cite{mahnob}. 

Finally, regarding evaluation metrics, we propose to not rely solely on accuracy because accuracy can misrepresent the actual performance of a model due to imbalanced class distributions. Thus, we argue in favor of additionally reporting F1-scores. Additionally, researchers should report standard deviations in order to provide readers with an estimate of model dispersion. All of these measures are implemented in the Logger block of \frameworkname{}. 

We believe that adhering to these guidelines can make research in the field of EEG-ER more transparent, replicable and comparable. Thus, our suggestions are in line with similar calls advocating more transparency and replicability in the broader field of affective computing~\cite{9597418}. As already discussed, datasets in the field of EEG-ER are small due to the effort involved in recording EEG data. Therefore, the field of EEG-ER might never be able to reach the level of standardization and comparability reached in other fields such as image recognition, face recognition or handwritten character recognition that can draw on large datasets with widely accepted ground-truth definitions and train-test splits such as ImageNet~\cite{RN790}, VGGFace2~\cite{RN713} or MNIST~\cite{RN981}. Nevertheless, we believe that establishing standards for the small datasets that exist in the field of EEG-ER will promote transparancy and comparability and thus accelerate progess in the field.

\section{Conclusion}

In our comprehensive review of papers published between 2018 and 2023, we critically examine papers focused on emotion recognition that implement machine learning and deep learning techniques on EEG. Our analysis highlights several inconsistencies regarding the evaluation pipeline that make direct comparisons between different studies challenging. Notable differences include varying evaluation metrics, inconsistent ground truth definition, and diverse train/test data splitting. A significant concern is the common use of a single dataset for evaluation, which limits the generalizability of the methods.
To improve the state of the art, we introduce a unified framework that can be used for a fair evaluation of new architectures and new datasets. This approach includes a range of pre-processing methods, the integration of prevalent EEG models and datasets (custom model and dataset usage is also available) and a module for the results evaluation. We provide a comprehensive set of evaluation metrics that facilitates the effective comparison of the results. 
We believe that our framework, %
equips the scientific community with solid benchmarks and a uniform comparison methodology, advancing consistent research in the field of emotion recognition from EEG.

\printbibliography %

\end{document}